\documentclass[useAMS,usenatbib]{mn2e}
\usepackage{graphicx}

\usepackage{ulem}

\title{AN ALTERNATIVE SYMBIOTIC CHANNEL TO TYPE Ia SUPERNOVAE}
\author[ L\"{u} et al. ]{Guoliang L\"{u}$^{1, 2}$\thanks{E-mail:
guolianglv@gmail.com (LGL)}, Chunhua Zhu$^{2, 3}$, Zhaojun Wang$^{2, 3}$, Na Wang$^1$\\
$^{1}$National Astronomical Observatories / Urumqi Observatory, the
Chinese Academy of Sciences, Urumqi, 830011, China\\
$^2$School of Physics, Xinjiang University, Urumqi, 830046,
China\\
$^{3}$School of Science, Xi'an Jiaotong University, Xi'an, 710049,
China }
\begin{document}

\date{}

\pagerange{\pageref{firstpage}--\pageref{lastpage}} \pubyear{2008}

\maketitle

\label{firstpage}

\begin{abstract}
By assuming an aspherical stellar wind with an equatorial disk from
a red giant, we investigate the production of Type Ia supernovae
(SNe Ia) via symbiotic channel. We estimate that the Galactic
birthrate of SNe Ia via symbiotic channel is between $1.03\times
10^{-3}$ and $2.27\times 10^{-5}$ yr$^{-1}$, the delay time of SNe
Ia has wide range from $\sim$ 0.07 to 5 Gyr. The results are greatly
affected by the outflow velocity and mass-loss rate of the
equatorial disk. Using our model, we discuss the progenitors of SN
2002ic and SN 2006X.

\end{abstract}

\section{Introduction}
Type Ia supernovae (SNe Ia) are exploding stars which are good
cosmological distance indicators and have been used to measure the
accelerated expansion of the Universe \citep{rie98,per99}. It is
widely accepted that progenitors of SNe Ia are mass accreting
carbon-oxygen (CO) white dwarfs (WDs) and they explode as a SN Ia
when their masses reach approximately the Chandrasekhar mass
\citep{nom84}. Two families of progenitor models have been proposed:
the double-degenerate model and the single-degenerate model. For the
double degenerate model, previous works indicated that the expected
accretion rates may cause the accretion-induced collapse of the CO
WDs and the formation of neutron stars \citep{nom85,sai85}. In the
single-degenerate model the mass donor is a main sequence (MS)
referred to as the WD+MS channel or a red giant (RG) referred to as
WD+RG channel \citep{van92,li97}. The latter is also called a
symbiotic channel.

According to \cite{han04} and \cite{men08}, the birth rate of SNe Ia
via the WD+MS channel can only account for up to 1/3 of
$3-4\times10^{-3}$yr$^{-1}$ observationally estimated by
\cite{van91} and \cite{cap97}. Other channels should be important.
Recently, \cite{wan09} showed that helium star donor channel is
noteworthy for producing SNe Ia ($\sim 1.2\times 10^{-3}$
yr$^{-1}$). However, helium associated with SNe Ia fails to be
detected \citep{mar03,mat05}, and \cite{kat08} suggested that SNe Ia
from helium star donor channel are rare. Therefore, symbiotic
channel is a possible candidate. By detecting Na I absorption lines
with low expansion velocities, \cite{pat07} suggested that the
companion of the progenitor of the SN 2006X may be an early RG
(However, \cite{chu08} showed that the absorption lines detected in
SN 2006X cannot form in the RG wind). \cite{vos08} studied the
pre-supernova archival X-ray images at the position of the recent SN
2007on, and they considered that its progenitor may be a symbiotic
binary. Unfortunately, a WD+RG binary system usually undergoes a
common envelope phase when RG overflows its Roche lobe in popular
theoretical model. WD+RG binaries are unlikely to become a main
channel for SNe Ia. \cite{yun95}, \cite{yun98}, \cite{han04} and
\cite{lu06} showed that the birthrate of SNe Ia via symbiotic
channel are much lower than that from WD+MS channel. In order to
stabilize the mass transfer process and avoid the common envelope,
\cite{hac99} assumed a mass-stripping model in which a wind from the
WDs strips mass from the RG. They obtained a high birth rate ($\sim
0.002$ yr$^{-1}$) of SNe Ia coming from WD+RG channel.

Most of previous theoretical works assumed that the cool giant in
symbiotic stars shares the same mass-loss rate with field giant and
the wind from the cool giant is spherical \citep{yun95,lu06,lu08}.
However, based on cm and mm/submm radio observations
\citep{sea93,mik03} and IRAS data \citep{ken88}, mass-loss rates for
the symbiotic giants are systematically higher than those reported
for field giants. Recently, \cite{zam08} found that the rotational
velocities of the giants in symbiotic stars are 1.5---4 times faster
than those of field giants. Using the relation between rotational
velocity and mass-loss rate found by \cite{nie88}, \cite{zam08}
estimated that the mass-loss rates of the symbiotic giants are
3---30 times higher than those of field giants. In addition,
\cite{obr06} suggested that the bipolarity in the 2006 outburst of
the recurrent nova RS Oph may result from an equatorial enhancement
in its cool giant. If the cool giant in symbiotic star has high
mass-loss rate and an aspherical stellar wind, the contribution of
symbiotic channel to total SNe Ia may be significantly enhanced.


In this work, assuming an aspherical stellar wind with equatorial
disk around the symbiotic giant, we show an alternative symbiotic
channel to SNe Ia. In $\S$ 2 we present our assumptions and describe
some details of the modeling algorithm. In $\S$ 3 we show WD+RG
systems in which SNe Ia are expected. The population synthesis'
 results are given in $\S$ 4. Main conclusions are in $\S$ 5.

\section{Progenitor Model}
For the binary evolution, we use a rapid binary star evolution code
of \cite{hur02}. When the primary has become a CO WD and the
secondary just evolves into first giant branch (FGB) or asymptotic
giant branch (AGB), an aspherical wind from the secondary is
considered. In other phases of binary evolution, we adopt the
descriptions of \cite{hur02}.
\subsection{Aspherical winds from cool giants in symbiotic stars}
\label{sec:asphe} In general, the stellar wind from a normal RG is
expected to be largely spherical due to the spherical stellar
surface and isotropic radiation. However, the majority ($>80\%$) of
observed planetary nebulae are found to have aspherical morphologies
\citep{zuc86,bal87}. This property can be explained by two models:
(i)the generalized wind-blown bubble in which a fast tenuous wind is
blown into a previously ejected slow wind (see a review by
\cite{fra99}); (ii)the interaction of the slow wind blown by an AGB
star with a collimated fast wind blown by its companion
\citep{sok00}. In two models, the slow wind is aspherically
distributed and the densest in the equatorial plane. The aspherical
wind with an equatorial disk may result from the stellar rotation
\citep{bjo93} or a simple dipole magnetic field \citep{mat00}.
\cite{asi95} showed that an anisotropic wind from AGB star can be
caused by combining rotation effects with the existence of an
inflated atmosphere formed by the stellar pulsation. Therefore, the
fast rotation and strong magnetic field are crucial physical
conditions for an equatorial disk.

An isolated RG usually has a low rotational velocity so that its
effect can be neglected. Furthermore, using a traditional method
applied for the solar magnetic field, \cite{sok92} estimated the
magnetic activity of AGB stars, and found that the level of activity
expected from single AGB stars is too low to explain the aspherical
wind. However, the situations in symbiotic stars are different.
\cite{sok02} showed that the cool companions in symbiotic systems
are likely to rotate much faster than isolated RGs due to accretion,
tidal interaction and back-flowing material. According to
measurements of the projected rotational velocities of the cool
giants 9 symbiotic stars and 28 field giants, \cite{zam08} found
that the rotational velocities of the giants in symbiotic stars are
1.5-4 times faster than those of field giants, which confirmed the
results of \cite{sok02}. Stellar magnetic activity is closely
related to the rotation. The cool giants with fast rotational
velocities are prime candidates for possessing strong magnetic field
\citep{sok02}. Therefore, the cool giants in symbiotic stars, having
the fast rotation velocity and strong magnetic field, can result in
aspherical winds. However, in binary systems, the orbital motion,
the magnetic field and the gravitational influence of the companions
can complicate the structure of the outflow
\citep{sok94,sok97,fra01}. A detailed structure of the winds is
beyond the scope of this work. By assuming several parameters, we
construct a primary aspherical model to describe the morphology of
winds from cool giants in symbiotic stars.

No comprehensive theory of mass loss for RGs exists at present.
\cite{hur00} applied the mass loss laws of \cite{rei75} and
\cite{vas93} to describe the mass-loss rates of FGB and AGB stars,
respectively. As mentioned in Introduction, the mass-loss rates for
the symbiotic giants are systematically higher than those reported
for field giants. We use a free parameters $\zeta$ to represent the
enhanced times of mass-loss rates during FGB and AGB phases for the
symbiotic giants by
\begin{equation}
\dot{M}_{\rm L}=\zeta\dot{M}_{\rm RG} \label{eq:eml}
\end{equation}
where $\dot{M}_{\rm RG}$ is the mass-loss rate of RG in
\cite{hur00}.

In the present work, the winds from cool giants in symbiotic stars
flow out via two ways: an equatorial disk and a spherical wind. The
total mass-loss rate $\dot{M}_{\rm L}$ is represented by
\begin{equation}
\dot{M}_{\rm L}=\dot{M}^{\rm d}+\dot{M}^{\rm sph}
\end{equation}
where $\dot{M}^{\rm d}$ and $\dot{M}^{\rm sph}$ give the mass-loss
rates via the equatorial disk and the spherical wind, respectively.
We assume that the ratio of the mass-loss rate in the equatorial
disk to the total mass-loss rate is represented by a free parameter
$\eta$, that is,
\begin{equation}
\dot{M}_{\rm L}=\dot{M}^{\rm d}/\eta=\dot{M}^{\rm sph}/(1-\eta).
\label{eq:dtr}
\end{equation}
In this work we make different numerical simulations for a wide
range of $\zeta$ and $\eta$.

The mass-loss rates $\dot{M}_{\rm L}$, $\dot{M}^{\rm d}$ and
$\dot{M}^{\rm sph}$ are affected by the rotational velocities and
the magnetic activities of the RGs in symbiotic stars. The
rotational velocities and the magnetic activities depend on the
binary evolutionary history \citep{sok02}. In the present paper we
focus on the effects of the aspherical wind with an equatorial disk
on the formation of SNe Ia. We neglect the different rotational
velocities and magnetic activities which result from the different
binary evolutionary history and result in different $\dot{M}_{\rm
L}$, $\dot{M}^{\rm d}$ and $\dot{M}^{\rm sph}$. This work
overestimates the contribution of the symbiotic stars with long
orbital periods to SN Ia. Recently, \cite{sok08} suggested the
existence of an extended zone above the AGB star where parcels of
gas do not reach the escape velocity. In general, the binary with an
AGB star have a long orbital periods. The low outflow velocity of
stellar wind greatly increases the efficiency of the companion
accreting stellar wind. In our work, we do not consider the extended
zone above the AGB star, which results in an underestimate the
contribution of the symbiotic stars with long orbital periods to SN
Ia. Therefore, it is acceptable for a primary model to neglect the
binary evolutionary history and the extended zone above the AGB
star.

\subsection{Wind accreting}
In symbiotic stars WDs accrete a fraction of  the stellar winds from
cool companions. In the present paper a stellar wind is composed of
a spherical wind and an equatorial disk. For the former, according
to the classical accretion formula in \citet{bon44}, \cite{bof88}
gave the mean accretion rate in binaries by
\begin{equation}
\dot{M}^{{\rm sph}}_{\rm
a}=\frac{-1}{\sqrt{1-e^2}}\left(\frac{GM_{{\rm WD}}}{v^2_{\rm
w}}\right)^2\frac{\xi_{\rm w}}{2a^2}
\frac{1}{(1+v^2)^{3/2}}\dot{M}^{\rm sph}, \label{eq:bona}
\end{equation}
where $1\le\xi_{\rm w}\le2$ is a parameter (Following \cite{hur00},
$\xi_{\rm w}=\frac{3}{2}$), $v_{\rm w}$ is the wind velocity and
\begin{equation}
v^2=\frac{v^2_{{\rm orb}}}{v^2_{\rm w}},~~ v^2_{{\rm
orb}}=\frac{GM_{\rm t}}{a},
\end{equation}
where $a$ is the semi-major axis of the ellipse, $v_{{\rm orb}}$ is
the orbital velocity and total mass $M_{\rm t}=M_{{\rm hot}}+M_{{\rm
cool}}$. For the equatorial disk, we neglect the diffusion of
equatorial disk, that is, the disk thickness is constant.  The
accretion rate of WD is relative to the crossing area of its
gravitational radius in orbital plane, which is approximately given
by :
\begin{equation}
\dot{M}_{\rm a}^{\rm d}=\frac{R_{\rm G}v_{\rm orb}}{\pi av_{\rm
w}}\dot{M}^{\rm d} \label{eq:diska}
\end{equation}
where $R_{\rm G}=2GM_{\rm WD}/(v_{\rm w}^2+v_{\rm orb}^2)$ is the
gravitational radius of WD accretor. If $v_{\rm orb}>>v_{\rm w}$,
the accretion rate given by Eq. (\ref{eq:diska}) may be higher than
$\dot{M}^{\rm d}$. It is necessary that the WD does not accrete more
mass than that lost by the RG. So we enforced the condition
$\dot{M}_{\rm a}^{\rm d}\leq0.9\dot{M}^{\rm d}$. For $v_{\rm
orb}<<v_{\rm w}$, we requested $\dot{M}_{\rm a}^{\rm d}$ is not
lower than $R_{\rm g}/ (\pi a) \dot{M}^{\rm d}$. Eq.
(\ref{eq:diska}) overestimates the contribution of the symbiotic
stars with long orbital periods to SN Ia because we neglect the
diffusion of equatorial disk. Total mass accretion rate is
\begin{equation}
\dot{M}_{\rm a}=\dot{M}_{\rm a}^{\rm d}+\dot{M}_{\rm a}^{\rm sph}
\end{equation}

Based on Eqs. (\ref{eq:bona}) and (\ref{eq:diska}), accretion rate
depends strongly on the wind outflow velocity $v_{\rm w}$ which is
not readily determined. For the spherical winds, we adopt the
prescription in \cite{hur02}, and $v_{\rm w}=\sqrt{2\beta_{\rm
w}\frac{GM_{\rm cool}}{R_{\rm cool}}}$ where $\beta_{\rm w}=1/8$. In
general, the equatorial disk has a lower outflow velocity than the
spherical wind \citep{bjo93,asi95}, and the typical wind velocity of
field giant is between $\sim$ 5 and 30 km s$^{-1}$. Due to the
existence of an extended zone above the AGB star\citep{sok08}, the
outflow velocity of the equatorial disk can be very low. In this
work the outflow velocity of the equatorial disk $v_{\rm w}$ is
taken as 2, 5 and 10 km s$^{-1}$ in different numerical simulations.

\subsection{Mass transfer rate of Roche lobe overflow}
When a secondary overflows its Roche lobe, we assume that there is
no equatorial disk for the secondary. At this time the mass transfer
via the inner Lagrangian point (L$_1$) can be dynamically unstable
or stable. If the mass ratio of the components ($q=M_{{\rm
donor}}/M_{{\rm accretor}}$) at the onset of Roche lobe overflow is
larger than a certain critical value $q_{\rm c}$, the mass transfer
is dynamically unstable and results in the formation of a common
envelope.  The issue of the criterion for dynamically unstable Roche
lobe overflow $q_{\rm c}$ is still open. \citet{hje87} did a
detailed study of stability of mass transfer using polytropic
models. \citet{han01,han02} showed that $q_{\rm c}$ depends heavily
on the assumed mass-transfer efficiency. Recently, \cite{che08}
studied $q_{\rm c}$ for dynamically stable mass transfer from a
giant star to a main sequence companion. They found that $q_{\rm c}$
almost linearly increases with the amount of the mass and angular
momentum lost during mass transfer. In this work, for normal main
sequence stars $q_{\rm c}=3.0$ while $q_{\rm c}=4.0$ when the
secondary is in Hertzsprung gap \citep{hur02}. Base on the
polytropic models in \citet{hje87}, \cite{web88} gave $q_{\rm c}$
for red giants by
\begin{equation}
q_{\rm c}=0.362+\frac{1}{3\times(1-M_{\rm c}/M)}, \label{eq:qcrit}
\end{equation}
where $M_{\rm c}$ and $M$ are core mass and mass of the donor,
respectively. In our work, we adopt the $q_{\rm c}$ of \cite{web88}.

When $q<q_{\rm c}$, the binary system undergoes a stable Roche lobe
mass transfer and the mass transfer rate is calculated by
\begin{equation}
\dot{M}_{\rm L}=3.0\time 10^{-6}[\ln (R_{\rm d}/R_{\rm Ld})]^3
M_\odot {\rm yr}^{-1} \label{eq:mstr}
\end{equation}
where $R_{\rm d}$ and $R_{\rm Ld}$ are donor's radius and Roche lobe
radius, respectively. Details can be seen in \cite{hur02}.

When $q>q_{\rm c}$, the binaries in which the giants overflow the
Roche lobe immediately evolve to the common envelope phase, while
the binaries for the main sequence stars overflowing the Roche lobes
will undergo the dynamically stable mass transfer.

\subsection{Evolution of accreting WD}
Instead of calculating the effects of accretion on to the WD
explicitly, we adopt the prescription of \cite{hac99} for the growth
of the mass of a CO WD by accretion of hydrogen-rich material from
its companion (also see \cite{han04,men08}). If the mass accretion
rate $\dot{M}_{\rm a}$ exceeds a critical value, $\dot{M}_{\rm cr}$,
the accreted hydrogen burns steadily on the surface of the WD at the
rate of $\dot{M}_{\rm cr}$. The unprocessed matter is assumed to be
lost from the systems as an optically thick wind at a rate of
$\dot{M}_{\rm wind}=\dot{M}_{\rm a}-\dot{M}_{\rm cr}$ \citep{hac96}.
The critical mass-accretion rate is given by
\begin{equation}
\dot{M}_{\rm cr}=5.3\times10^{-7}\frac{1.7-X}{X}(M_{\rm
WD}-0.4)M_\odot {\rm yr}^{-1}
\end{equation}
where $X$ is the hydrogen mass fraction and is 0.7 in this work. If
the mass-accretion rate $\dot{M}_{\rm a}$ is less than $\dot{M}_{\rm
cr}$ but higher than $\dot{M}_{\rm st}=\frac{1}{2}\dot{M}_{\rm cr}$,
it is assumed that there is no mass loss and hydrogen-shell burning
is steady. If $\dot{M}_{\rm a}$ is between $\frac{1}{2}\dot{M}_{\rm
cr}$ and $\frac{1}{8}\dot{M}_{\rm cr}$, the accreting WD undergoes
very weak hydrogen-shell flashes, where we assume that the processed
mass can be retained. If $\dot{M}_{\rm a}$ is lower than
$\frac{1}{8}\dot{M}_{\rm cr}$, hydrogen-shell flashes are so strong
that no mass can be accumulated by the accreting WD. The growth rate
of the mass of the helium layer on top of the CO WD can be written
as
\begin{equation}
\dot{M}_{\rm He}=\eta_{\rm H}\dot{M}_{\rm a}
\end{equation}
where
\begin{equation}
\eta_{\rm H}=\left \{
\begin{array}{ll}
\dot{M}_{\rm cr}/\dot{M}_{\rm a}, & \dot{M}_{\rm a}>\dot{M}_{\rm
cr},\\
1, & \dot{M}_{\rm cr}\geq \dot{M}_{\rm a}\geq
\frac{1}{8}\dot{M}_{\rm cr},\\
0, &\dot{M}_{\rm a}< \frac{1}{8}\dot{M}_{\rm cr}.
\end{array}
\right.
\end{equation}

When the mass of the helium layer reaches a certain value, helium is
possible ignited. If helium-shell flashes occur, a part of the
envelope mass is assumed to be blown off. The mass accumulation
efficiency for helium-shell flashes,  $\eta_{\rm He}$, is given by
\cite{kat04}. Then, the mass growth rate of the CO WD, $\dot{M}_{\rm
WD}$, is
\begin{equation}
\dot{M}_{\rm WD}=\eta_{\rm H}\eta_{\rm He}\dot{M}_{\rm a}.
\end{equation}
When the mass of accreting CO WD reaches 1.378 $M_\odot$, it
explodes as a SN Ia.

\section{WD+RG systems in which SNe Ia are expected}
\begin{figure*}
\begin{tabular}{c}
\includegraphics[totalheight=6.8in,width=5.0in,angle=-90]{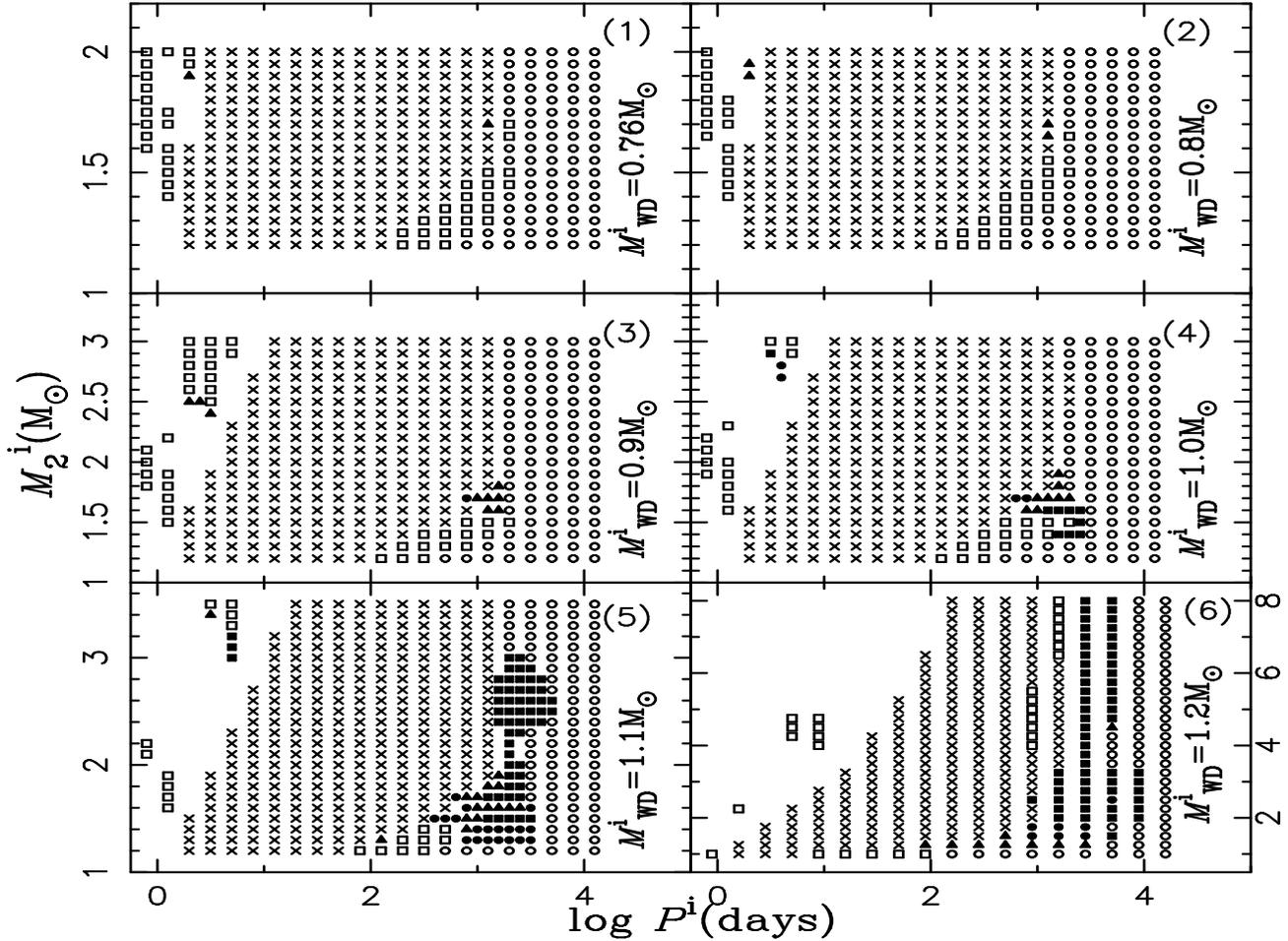}

\end{tabular}
\caption{---Final outcomes of the binary evolution in the initial
orbital period-secondary mass ($\log P^{\rm i}, M^{\rm i}_{2}$)
plane of the CO WD+RG binary, where $P^{\rm i}$ is the initial
orbital period and $M^{\rm i}_{2}$ is the initial mass of the donor
star (for different initial WD masses as indicated in each panel).
Filled squares indicate SN Ia explosions during an optically thick
wind phase $( \dot{M}_{\rm a}\geq \dot{M}_{\rm cr})$, filled circles
SN Ia explosions during stable hydrogen-shell burning ($\dot{M}_{\rm
cr} >  \dot{M}_{\rm a} \geq \frac{1}{2} \dot{M}_{\rm cr}$), filled
triangles SN Ia explosions during mildly unstable hydrogen-shell
burning ($\frac{1}{2}\dot{M}_{\rm cr} >  \dot{M}_{\rm a} \geq
\frac{1}{8} \dot{M}_{\rm cr}$). Crosses show binary systems which
undergo common envelope evolution during WD+RG phase, empty squares
represent binary systems which experience stable Roche lobe
overflows during WD+RG phase, empty circles give binary systems
which are detached systems during WD+RG phase. They can produce
symbiotic phenomena while they can not explode as SNe Ia. The
numbers along the right $y$ axis in panel 6 only are for $M^{\rm
i}_{\rm WD}=1.2 M_\odot$. Details are in text. } \label{fig:mprg}
\end{figure*}

In this section, according to the assumptions in the above section,
we simulate the evolutions of the binary systems with a CO WD and a
MS. All input parameters are the same with those in case 1, that is,
the outflow velocity of equatorial disk $v_{\rm w}=5$ km s$^{-1}$,
$\eta=0.9$ and $\zeta=10$. The initial masses of WDs are 0.76, 0.8,
0.9, 1.0, 1.1, and 1.2 $M_\odot$, respectively. The initial orbital
periods are from 0.8 to 11000 days with $\Delta \log P^{\rm i}=0.1$
days. The initial masses of MSs are between 1.2 and 2.0 $M_\odot$
with $\Delta M^{\rm i}_2=0.05 M_\odot$ when $M^{\rm i}_{\rm WD}=0.76
M_\odot$ and $M^{\rm i}_{\rm WD}=0.80 M_\odot$, between 1.15 and 3.0
$M_\odot$ with $\Delta M^{\rm i}_2=0.1 M_\odot$ when $M^{\rm i}_{\rm
WD}=0.9 M_\odot$ and $M^{\rm i}_{\rm WD}=1.0 M_\odot$, between 1.1
and 3.5 $M_\odot$ with $\Delta M^{\rm i}_2=0.1 M_\odot$ when $M^{\rm
i}_{\rm WD}=1.1 M_\odot$, between 1.0 and 8 $M_\odot$ with $\Delta
M^{\rm i}_2=0.25 M_\odot$ when $M^{\rm i}_{\rm WD}=1.2 M_\odot$,
respectively.

Fig. \ref{fig:mprg} shows WD+RG binary systems in the initial
orbital period-secondary masse ($\log P^{\rm i}, M^{\rm i}_{2}$)
plane. Filled symbols give the binary systems in which CO WDs
explode eventually as SNe Ia: Filled squares indicate SN Ia
explosions during an optically thick wind phase $( \dot{M}_{\rm
a}\geq \dot{M}_{\rm cr})$; filled circles SN Ia explosions during
stable hydrogen-shell burning ($\dot{M}_{\rm cr} >  \dot{M}_{\rm a}
\geq \frac{1}{2} \dot{M}_{\rm cr}$); filled triangles SN Ia
explosions during mildly unstable hydrogen-shell burning
($\frac{1}{2}\dot{M}_{\rm cr} >  \dot{M}_{\rm a} \geq \frac{1}{8}
\dot{M}_{\rm cr}$). Crosses show binary systems which undergo common
envelope evolution during WD+RG phases, empty squares represent
binary systems which experience stable Roche lobe overflows during
WD+RG phases, empty circles give binary systems which are detached
WD+RG systems. They can produce symbiotic phenomena while they can
not explode as SNe Ia. The binary systems which are not plotted can
not evolve to WD+RG phases. They either explode as SNe Ia via WD+MS
channel, or become WD+dwarf or helium MS systems.

According to Fig. \ref{fig:mprg}, the progenitors of SNe Ia via
symbiotic channel are mainly split into left and right regions. The
progenitors in the left region have short initial orbital periods.
Table \ref{tab:fgbsn} shows an example. The secondary overflows
Roche lobe during Hertzsprung gap.  Due to $M_{\rm donor}/M_{\rm WD}
< q_{\rm c}$, the progenitor experiences stable mass transfer.
During Roche lobe overflows, the great deal matter of the secondary
has been transferred to the CO WD so that the CO WD becomes more
massive while the secondary mass decreases. When the secondary
evolves to FGB phase, $M_{\rm donor}/M_{\rm WD}$ has been lower than
$q_{\rm c}$. Therefore the progenitor undergoes a stable mass
transfer in FGB phase till CO WD explodes as SN Ia. In the whole
process, there is no aspherical wind with equatorial disk from the
secondary. The progenitors in the right region have long initial
orbital periods so that the secondaries can evolve into AGB phase.
Table \ref{tab:agbsn} shows an example. Due to the high mass-loss
rate of RG and the high accretion rate of CO WD resulting from an
aspherical wind with equatorial disk, the ratio of mass of RG to
that of CO WD is lower than $q_{\rm c}$ when secondary overflows its
Roche lobe. Therefore, the binary system undergoes a stable matter
transfer until CO WD explodes as SN Ia. In the progenitors with
longer initial orbital periods, the secondaries never overflow Roche
lobe and the accretion of CO WDs mainly depends on the aspherical
wind with equatorial disk.

Compared with the previous work \citep{hac99,han04,men08}, the
progenitors in the present work have longer initial orbital periods
(up to $\sim$ 10000 days) and wider ranges of initial masses of
secondaries (from $\sim 1.0$ to 8 $M_\odot$).

\begin{table}
 \begin{minipage}{80mm}
  \caption{An example for the progenitor of SN Ia with a short initial orbital
           period. The initial mass of the CO WD is 0.8 $M_\odot$, the initial
           mass of the secondary is 1.9 $M_\odot$ and the initial orbital period is 1.99 days.
           The first column gives the evolutionary age. Columns 2, and 3 show the masses of CO WD
           and the secondary, respectively. The letters in parentheses of column 3 mean
           the evolutionary phases of the secondary.
           MS represents the main sequence, HG for Hertzsprung gap.
           The forth column
           shows the orbital period. Column 5 gives the ratio of the secondary radius
           to its Roche lobe radius. The last column shows the critical mass ratio $q_{\rm c}$.
           All input physical parameters are same with those in case 1.}
  \tabcolsep0.8mm
  \begin{tabular}{|lllllll|}
  \hline\hline
AGE ($10^6$yr) & $M_{\rm WD} (M_{\odot})$&$M_{2} (M_\odot)$&$P$
(Days)&$R_{2}/R_{\rm L}$&$q_{\rm
c}$\\
0.0000&0.80&1.90(MS)  &1.99 & 0.373&---\\
1352.0&0.80&1.90(HG)  &1.99 & 0.813&---\\
1357.6&0.80&1.90(HG)  &1.99 & 1.001& 4\\
1363.8&1.26&0.70(FGB) &2.21 & 2.229& 0.83 \\
1364.1&$\sim$1.378&0.55 (FGB)&3.25 & 2.003& 0.86\\

\hline
 \label{tab:fgbsn}
\end{tabular}
\end{minipage}
\end{table}
\begin{table}
 \begin{minipage}{80mm}
  \caption{Similar to Table \ref{tab:fgbsn}, but for the progenitor of SN Ia with a long initial orbital
           period. CHeB means core helium burning. The initial mass of the CO WD is 0.8 $M_\odot$, the initial
           mass of the secondary is 1.7 $M_\odot$ and the initial orbital period is 1256 days.
           }
  \tabcolsep0.8mm
  \begin{tabular}{|lllllll|}
  \hline\hline
AGE ($10^6$yr) & $M_{\rm WD} (M_{\odot})$&$M_{2} (M_\odot)$&$P$
(Days)&$R_{2}/R_{\rm L}$&$q_{\rm
c}$\\
0.0000&0.80&1.70(MS)  &1256 & 0.005&---\\
1873.6&0.80&1.70(HG)  &1256 & 0.010&---\\
1901.1&0.80&1.70(FGB) &1256 & 0.014&---\\
1977.1&0.98&1.33(CHeB)&795  & 0.064&---\\
2108.6&0.98&1.30(AGB) &816  & 0.125&---\\
2112.3&1.10&1.05(AGB) &666  & 1.001&1.039\\
2113.5&$\sim$1.378&0.55(AGB)&890&1.715&4.64\\

\hline
 \label{tab:agbsn}
\end{tabular}
\end{minipage}
\end{table}

\section{Population Synthesis}
We construct a set of models in which we vary different input
parameters relevant to SNe Ia produced by aspherical stellar wind.
Table \ref{tab:cases} lists all cases considered in the present
work.

\begin{table}
 \begin{minipage}{80mm}
  \caption{Parameters of the models for SNe Ia via symbiotic channel.
           The parameter $v_{\rm w}$ in the second column is the
           outflow velocity of equatorial disk. The parameter $\eta$ in
           column three gives the ratio of the mass-loss rate in
           equatorial disk to the total mass-loss rate [See Eq. (\ref{eq:dtr})].
           The parameter $\zeta$ in the fourth column is the
           enhanced times of the mass-loss rate during FGB and AGB
           phases for the secondaries [See Eq. (\ref{eq:eml})].}
  \tabcolsep5.0mm
  \begin{tabular}{|llll|}
  \hline\hline
Cases & $v_{\rm w}$ (km s$^{-1}$)&$\eta$&$\zeta$\\
Case 1 & 5 & 0.9 & 10\\
Case 2 & 2 & 0.9 & 10\\
Case 3 & 10& 0.9 & 10\\
Case 4 & 5 & 0.75& 10\\
Case 5 & 5 & 0.5 & 10\\
Case 6 & 5 & 0.25& 10\\
Case 7 & 5 & 0.9 & 5 \\
Case 8 & 5 & 0.9 & 30\\
\hline
 \label{tab:cases}
\end{tabular}
\end{minipage}
\end{table}
\subsection{Parameters for binary evolution and binary population synthesis}
In order to investigate the birth rate of SNe Ia, we carry out
binary population synthesis via Monte Carlo simulation technique.
Binary evolution is affected by some uncertain input parameters.  In
this work, the rapid binary star evolution code of \cite{hur02} is
used. If any input parameter is not specially mentioned it is taken
as default value in \cite{hur02}. The metallicity $Z$=0.02 is
adopted. We assume that all binaries have initially circular orbits,
and we follow the evolution of both components by the rapid binary
evolution code, including the effect of tides on binary evolution
\citep{hur02}.

For the population synthesis of binary stars, the main input model
parameters are: (i) the initial mass function (IMF) of the
primaries; (ii) the mass-ratio distribution of the binaries; (iii)
 the distribution of orbital separations.

A simple approximation to the IMF of \citet{mil79} is used. The
primary mass is generated using  the formula suggested by
\citet{egg89}
\begin{equation}
M_1=\frac{0.19X}{(1-X)^{0.75}+0.032(1-X)^{0.25}},
\end{equation}
where $X$ is a random variable uniformly distributed in the range
[0,1],  and $M_1$ is the primary mass from $0.8M_\odot$ to
$8M_\odot$.

For the mass-ratio distribution of binary systems, we consider only
a constant distribution \citep{maz92},
\begin{equation}
n(q)=1,~~    0< q \leq 1,
\end{equation}
where $q=M_2/M_1$.

The distribution of separations is given by
\begin{equation}
\log a =5X+1,
\end{equation}
where $X$ is a random variable uniformly distributed in the range
[0,1] and $a$ is in $R_\odot$.

In order to investigate the birthrates of SNe Ia, we assume simply a
constant star formation rate over last 15 Gyr\citep{han95}, or a
single starburst\citep{han04,zha05}. In the case of a constant star
formation rate, we assume that a binary with its primary more
massive than 0.8 $M_\odot$ is formed
annually\citep{phi89,yun93,han95}. In the case of a single star
burst we assume a burst producing $2\times 10^7$ binary systems in
which the primaries are more massive than 0.8 $M_\odot$, which gives
a statistical error for our Monte Carlo simulation lower than 5 per
cent for SNe Ia via symbiotic channel in every case in Table
\ref{tab:cases}.

\subsection{Birthrate of SNe Ia via symbiotic channel}
\begin{figure}
\begin{tabular}{c}
\includegraphics[totalheight=3.2in,width=3.0in,angle=-90]{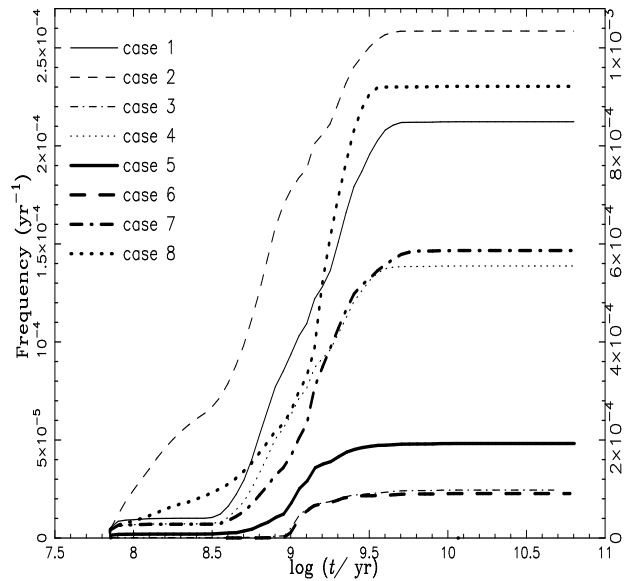}

\end{tabular}
\caption{---The evolution of birthrates of SNe Ia for a constant
star-formation rate. The key to the line-styles representing
different cases is given in the upper left corner. The numbers along
the right $y$ axis are only for case 2.
            }
\label{fig:galab}
\end{figure}

Fig. \ref{fig:galab} shows the Galactic birthrates of SNe Ia via
symbiotic channel. They are between $1.03\times 10^{-3}$ (case 2)
and $2.27\times 10^{-5} {\rm yr^{-1}}$ (case 6). Different outflow
velocities ($v_{\rm w}=2, 5, 10 {\rm km s^{-1}}$) of equatorial disk
in cases 1, 2 and 3 have a great effects on the birthrates with a
factor of $\sim$ 40. Different ratios of the mass-loss rate in the
equatorial disk to total mass-loss rate in cases 1, 4, 5 and 6,
$\eta$s ($\eta=0.9, 0.75, 0.5, 0.25$), give uncertainty of the
birthrates within a factor of $\sim$ 9. Different enhanced times of
the mass-loss rate for the symbiotic giants in cases 1, 7 and 8,
$\zeta$ ($\zeta=10, 5, 30$), have a weak effect within a factor of
1.6. The observationally estimated the Galactic birthrate of SNe Ia
by \cite{van91} and \cite{cap97} is $3-4\times10^{-3}$yr$^{-1}$. In
cases 3, 5 and 6, the contribution of SNe Ia via symbiotic channel
to total SNe Ia is negligible. However, in case 2 with low outflow
velocities ($v_{\rm w} = 2\ {\rm km\ s^{-1}}$) and high $\eta$
($\eta = 0.90$), the contribution is approximately 1/3, which is
comparable with that via WD+MS channel in \cite{han04}. Our results
are greatly affected by the low outflow velocity of equatorial disk
and its mass-loss rate.

\begin{figure}
\begin{tabular}{c}
\includegraphics[totalheight=3.2in,width=3.0in,angle=-90]{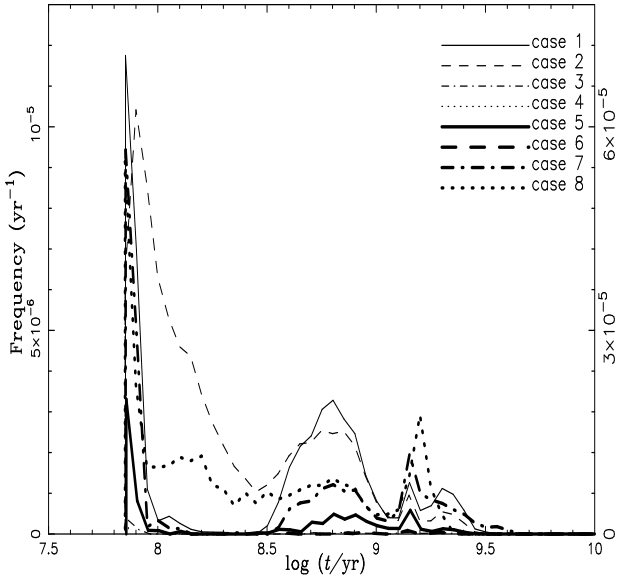}

\end{tabular}
\caption{---Similar to Fig. \ref{fig:galab}, but for a single star
burst of $2\times10^7$ binary systems. The numbers along the right
$y$ axis are only for case 2.
            }
\label{fig:sinb}
\end{figure}

Fig. \ref{fig:sinb} displays the evolution of birthrates of SNe Ia
for a single star burst of $2\times 10^{7}$ binary systems. Using
observations of the evolution of SNe Ia rate with redshift,
\cite{man06} suggested that SNe Ia have a wide range of delay time
from $< 0.1$ Gyr to $> 10$ Gyr. The delay time is defined as the age
at the explosion of the SN Ia progenitor from its birth. In the
single degenerate scenario, the delay time is closely related to the
secondary lifetime and thus the initial mass of the secondary
$M^{\rm i}_{2}$ \citep{cheli07,hac08b}. In WD+MS channel,
\cite{li97}, \cite{han04}, and \cite{men08} showed that $M^{\rm
i}_{2}$ is between 2 and 3.5 $M_\odot$, which indicates that the
range of the delay time is from $\sim$ 0.1 Gyr to 1 Gyr.
In order to obtain a wide delay time from $\sim$ 0.1 to 10 Gyr,
\cite{hac08b} assumed that optically thick winds from the
mass-accreting CO WD and mass-stripping from the companion star by
the WD wind. Assuming an aspherical stellar wind with an equatorial
disk from cool giants in symbiotic stars, we give a very wide delay
time range from $\sim$ 0.07 Gyr to 5 Gyr. SNe Ia with shorter delay
time than 0.1 Gyr result from those progenitors with more massive
$M^{\rm i}_{2}$ than $\sim 3.5 M_\odot$, SNe Ia with longer delay
time than 1 Gyr from those progenitors with lower $M^{\rm i}_{2}$
than 2$M_\odot$. Because the delay time determined greatly by the
initial mass of the secondary, Fig. \ref{fig:sinb} can be easily
explained by  the distribution of the initial masses of the
secondaries for SNe Ia in the next subsection.




\subsection{Distribution of initial parameters}
\begin{figure}
\begin{tabular}{c}
\includegraphics[totalheight=3.2in,width=3.0in,angle=-90]{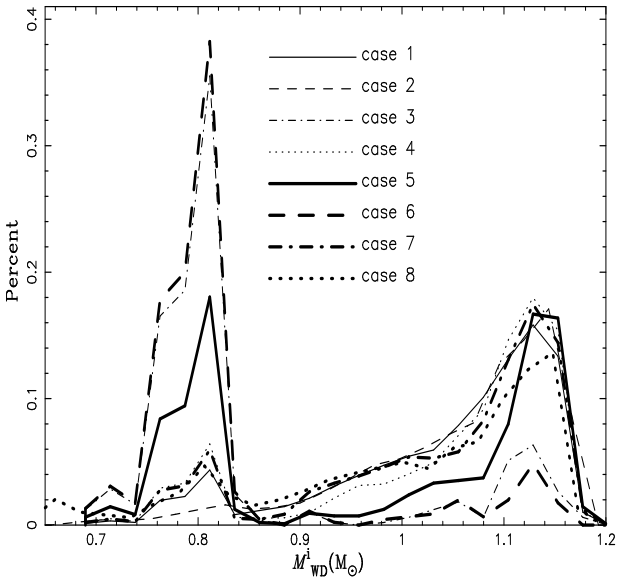}

\end{tabular}
\caption{---The distribution of the initial masses of the CO WDs for
the progenitors of SNe Ia (the number in every case is normalized to
1 and the width of the bin is 0.03 $M_\odot$). The cases are
indicated in the middle top.
            }
\label{fig:miwd}
\end{figure}

Fig. \ref{fig:miwd} shows the distribution of the initial masses of
the CO WDs that produce ultimately a SN Ia according to our models.
There are obviously two peaks. The left peak is at about 0.8
$M_\odot$, and results mainly from the binary systems with short
initial periods like that showed in Table \ref{tab:fgbsn} and with
long initial periods like that showed in Table \ref{tab:agbsn}. The
right peak is at about 1.1 $M_\odot$, and mainly results from binary
systems with long initial
periods.

\begin{figure}
\begin{tabular}{c}
\includegraphics[totalheight=3.2in,width=3.0in,angle=-90]{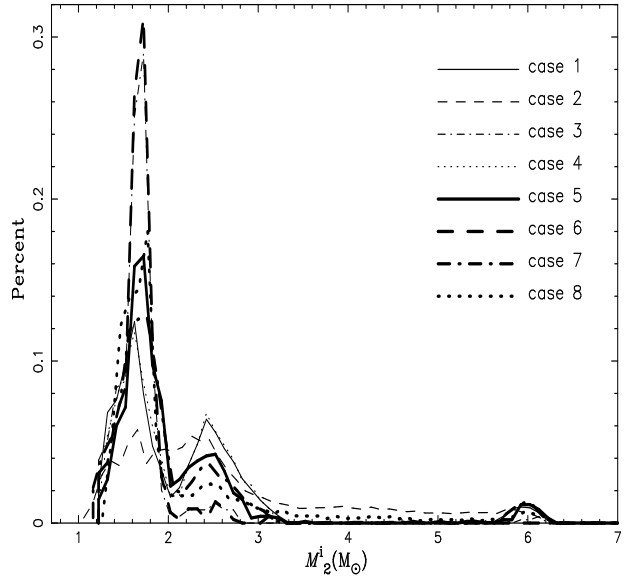}

\end{tabular}
\caption{---The distribution of the initial masses of the
secondaries for the progenitors of SNe Ia (the number in every case
is normalized to 1 and the width of the bin is 0.3 $M_\odot$). The
cases are indicated in the right top.
            }
\label{fig:mirg}
\end{figure}

\begin{figure}
\begin{tabular}{c}
\includegraphics[totalheight=3.2in,width=3.0in,angle=-90]{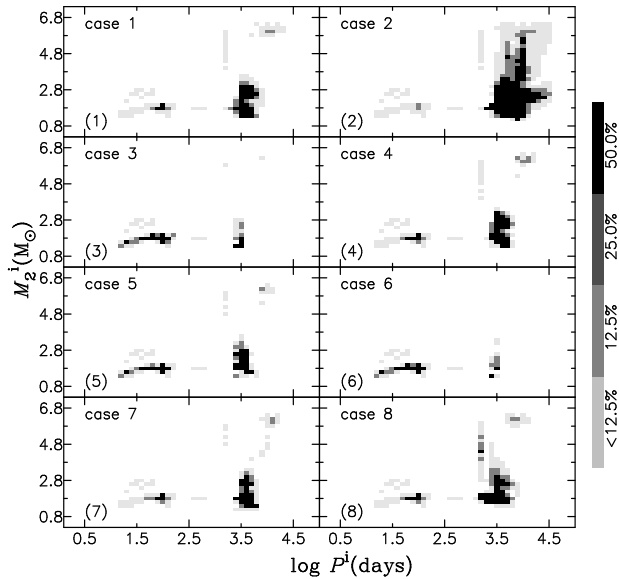}
\end{tabular}
\caption{---Gray-scale maps of secondary initial
            masses $M^{\rm i}_{\rm 2}$ vs. initial orbital period $P^{\rm i}$ distribution
            for the progenitors of SNe Ia. The gradations of gray-scale
            correspond to the regions where the number density of systems is,
            respectively,  within 1 -- 1/2,
            1/2 -- 1/4, 1/4 -- 1/8, 1/8 -- 0 of the maximum of
             ${{{\partial^2{N}}\over{\partial {\log P_i}}{\partial {\log
            M_i}}}}$, and blank regions do not contain any stars.
           The cases shown in particular panels are indicated in their left-up corners.
            }
\label{fig:mipi}
\end{figure}

Fig. \ref{fig:mirg} gives the distribution of the initial masses of
the secondaries for SNe Ia. The distribution shows three regions.
The left region is between $\sim 1.2 M_\odot$ ($1.0 M_\odot$ for
case 2) and $\sim 2.0 M_\odot$. These SNe Ia have delay time longer
than about 1 Gyr and correspond to the right peak of Fig.
\ref{fig:sinb}. The middle region is from $\sim 2.0 M_\odot$ to $3.0
M_\odot$, and results from the progenitors with long orbital
periods. These SNe Ia have delay time between $\sim$ 0.5 and 1.0 Gyr
and correspond to the middle peak of Fig. \ref{fig:sinb}. The right
region is more massive mass than $\sim 3.0 M_\odot$. These SNe Ia
have very short delay time and correspond to the left peak of Fig.
\ref{fig:sinb}. Except case 6, there is a peak at about $6.0
M_\odot$ (also see Fig. \ref{fig:mipi}). The peak results from the
binaries in which primary initial masses ($M^{\rm i}_1$) are more
massive than $\sim 5.8 M_\odot$, secondary initial masses are
between $\sim 5.8$ and $M^{\rm i}_1$, and initial orbital periods
are about 10000 days. For these binaries, before the primaries
overflow their Roche lobe during AGB phase, the secondaries can
accrete some material so that the ratio of primary mass to secondary
mass is lower than the critical value $q_{\rm c}$. These binaries
avoid the common envelope evolution. Then, they have wide binary
separations so that the secondaries can evolve RGs and form
aspherical stellar winds.


Fig. \ref{fig:mipi} shows the distributions of the progenitors of
SNe Ia, in the ``initial secondary mass -- initial orbital period''
plane. According to Fig. \ref{fig:mipi}, the distribution of the
initial orbital periods should have double peaks. The evolution of
progenitors with short and long orbital periods can be explained by
Tables \ref{tab:fgbsn} and \ref{tab:agbsn}, respectively.
\cite{men08} showed that the distribution of progenitors of SNe Ia
via WD+MS channel is between $\sim$ 1 days and 10 days. The orbital
distribution via symbiotic channel in this work is much longer than
their distribution.

\subsection{Progenitor of SN 2002ic}
\label{sec:2002ic} SN 2002ic was the first SN Ia for which
circumstellar (CS) hydrogen has been detected unambiguously
\citep{ham03}. The evolutionary origin of SN 2002ic has been
investigated by \cite{liv03} on the basis of a common envelope
evolution model, by \cite{han06} on the basis of the delayed
dynamical instability model of binary mass transfer, by
\cite{wood06} on the basis of a recurrent nova model with a RG, by
\cite{hac08a} on the efficient mass-stripping from the companion
star by the WD wind. Spectropolarimetry observations suggest that SN
2002ic exploded inside a dense, clumpy CS environment, quite
possibly with a disk-like geometry \citep{wan04}. Subaru
spectroscopic observations also provide evidence for an interaction
of SN ejecta with a hydrogen-rich aspherical CS medium
\citep{den04}.
\begin{figure}
\begin{tabular}{c}
\includegraphics[totalheight=3.2in,width=3.0in,angle=-90]{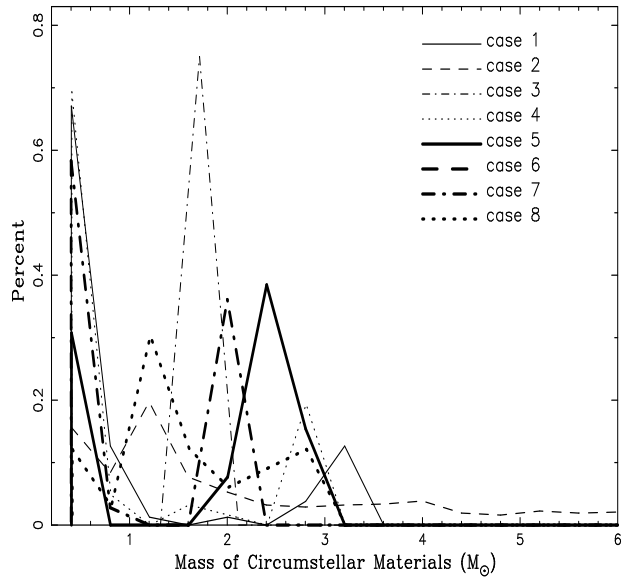}
\end{tabular}
\caption{---The distribution of total masses of CS material for
            SN 2002ic(the number in every case
            is normalized to 1 and the width of the bin is 0.4 $M_\odot$).
            The total masses of CS material is give by Eq. (\ref{eq:mcs}) in which
            $D$ is $1\times10^{17}$ cm.
            }
\label{fig:cirm}
\end{figure}

According to the above descriptions, the progenitor of SN 2002ic
should have a large amount of CS material which has a disk-like
geometry. In our model, the mass of the CS material is approximately
calculated by
\begin{equation}
M_{\rm CS}\approx\frac{D}{v^{\rm d}_{\rm w}}\eta \dot{M}_{\rm
L}+\frac{D}{v^{\rm s}_{\rm w}}(1-\eta)\dot{M}_{\rm L}-\Delta M_{\rm
WD} \label{eq:mcs}
\end{equation}
where $D$ is a distance from SN Ia, $\Delta M_{\rm WD}$ is the
increasing mass of WD for a span of $\frac{D}{v^{\rm d}_{\rm w}}$,
$v^{\rm d}_{\rm w}$ and $v^{\rm s}_{\rm w}$ are the outflow velocity
of the equatorial disk and spherical stellar wind, respectively. By
assuming that SN Ia explodes inside a spherically symmetric CS
envelope, \cite{chu04} estimated the total mass of the CS material
for SN 2002ic is about 0.4 $M_\odot$ within a radius of about
$7\times10^{15}$ cm, and for SN 1997cy about 6 $M_\odot$ within a
radius of about $2\times10^{16}$ cm. \cite{wan04} suggested that
there are several solar masses of material asymmetrically arrayed to
distances of $\sim 3\times 10^{17}$ cm. Therefore, we select the
progenitor of SN 2002ic by two conditions: (i)the mass of CS
material calculated by Eq. (\ref{eq:mcs}) being larger than $0.4
M_\odot$ within a radius of $10^{17}$ cm; (ii)a CS with a disk-like
geometry which can interact with the ejecta of SN 2002ic. In our
model there is always an equatorial disk around the secondary before
it overflows its Roche lobe.

\begin{figure}
\begin{tabular}{c}
\includegraphics[totalheight=3.2in,width=3.0in,angle=-90]{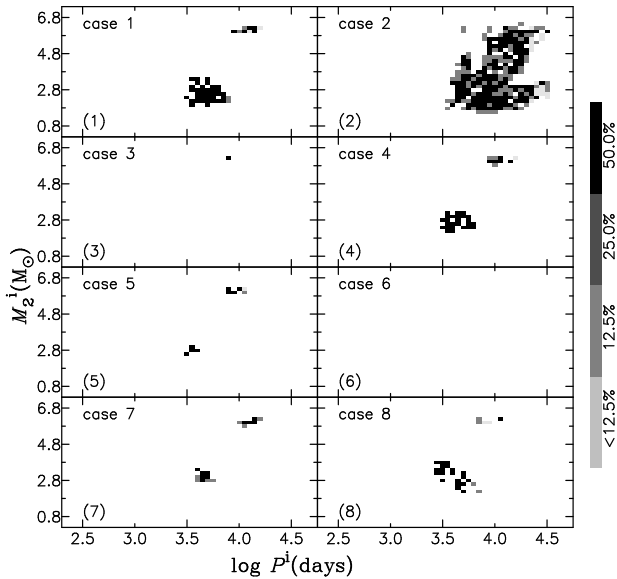}
\end{tabular}
\caption{---Similar to Fig. \ref{fig:mipi}, but for Gray-scale maps
            of initial secondary masses $M_{\rm 2}$ vs. initial orbital period $P$ distribution
            for the progenitors of SN 2002ic.
            }
\label{fig:mp02}
\end{figure}

In Fig. \ref{fig:cirm}, we give the distribution of total masses of
CS material for SNe Ia like 2002ic within a radius of $10^{17}$ cm.
The majority of the CS material lies in the orbital plane. Fig.
\ref{fig:mp02} shows gray-scale maps of initial secondary masses
$M_{\rm 2}$ vs. initial orbital period $P$ distribution for the
progenitors of SN 2002ic. In our model, the mass of CS material
depends on the initial secondary mass. For examples, there are two
peaks in Fig. \ref{fig:cirm} and two regions in Fig. \ref{fig:mp02}
for case 1. The left peak around $\sim 0.6 M_\odot$ in Fig.
\ref{fig:cirm} originates from the low region in Fig.
\ref{fig:mp02}, and the right peak around $\sim 0.6 M_\odot$ in Fig.
\ref{fig:cirm} from the up region in Fig. \ref{fig:mp02}.

\begin{figure}
\begin{tabular}{c}
\includegraphics[totalheight=3.2in,width=2.3in,angle=-90]{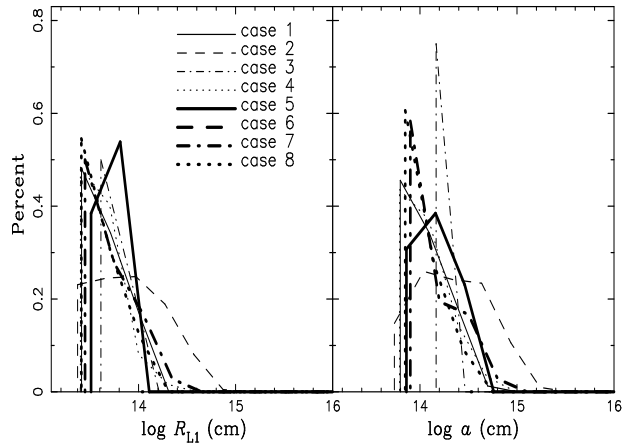}
\end{tabular}
\caption{---Distributions of the Roche lobe radius of the accreting
WD ($R_{\rm L1}$) and the binary separation of the progenitors of SN
2002ic prior to their explosion as a SN Ia. The left and right
panels are for Roche lobe radii of the accreting WDs and the binary
separations, respectively.
            }
\label{fig:sep02}
\end{figure}

SN 2002ic appeared to be a normal SN Ia from $\sim$ 5 to 20 days
after explosion \citep{wood02,ham03}, and brightened to twice the
luminosity of a normal SN Ia and showed strong H$_{\alpha}$ emission
around 22 days after explosion \citep{ham03}. The standard
brightness during the first 20 days after explosion and the
suddenness of the brightness around 22 days implied that the
original SN explosion expanded into a region with little CS
material, and then encountered a region with significant CS material
\citep{wood06}. Investigating the high-resolution optical
spectroscopy at 256 d and HK-band infrared photometry at +278 and
+380 d, \cite{kot04} suggested that there is a dense and slow-moving
($\sim$ 100 km s$^{-1}$) outflow, and a dusty CS material in SN
2002ic. In our model, a cavity with little CS material around the
accreting WD results from the WD's accretion, the dense and
slow-moving outflow originates from the dense equatorial disk which
have collided with the ejecta from SN explosion, and dust in CS
material is formed in the dense equatorial disk. The radius of the
cavity is roughly equal to the Roche lobe radius of the accreting
WD, $R_{\rm L1}$. The region out of the cavity has large amount of
CS material which mainly originates from the equatorial disk. Due to
orbital movement, the CS material out of the cavity has complicated
structure. However, the most dense region lies on the equatorial
disk around the secondary. The region where the ejecta from SN
2002ic firstly encountered the significant CS material is between
$R_{\rm L1}$ and binary separation. Fig. \ref{fig:sep02} shows the
distributions of $R_{\rm L1}$ and the binary separation of the
progenitors of SN 2002ic prior to their explosion as a SN Ia. The
observationally estimated region is $\sim2\times10^{15}$ cm away
from the explosion. Obviously, the majority of binary separations in
our work are shorter than $2\times10^{15}$ cm. A few binary
separations in case 2 in which the equatorial disk has a low outflow
velocity are longer than $2\times10^{15}$ cm. As mentioned in \S
\ref{sec:asphe}, we do not consider an extended zone above the AGB
star, which results in the underestimate for contribution of the
symbiotic stars with long orbital periods to SN Ia.

We calculate the birthrates of SNe Ia like SN 2002ic in the Galaxy.
From cases 1 to 8, they are $4.0\times 10^{-6}, 3.1\times 10^{-5},
2.0\times10^{-7}, 3.1\times 10^{-6}, 6.5\times 10^{-7}, \sim 0.0,
1.8\times 10^{-6}$ and $1.6\times 10^{-6} {\rm yr^{-1}}$,
respectively. Compared with $3-4\times10^{-3}$ yr$^{-1}$ which is
the birthrates of SNe Ia observationally estimated in the Galaxy,
the birthrates of SNe Ia like SN 2002ic in this work are not more
than 1\% of the total birthrates of SNe Ia. SN 2002ic is very rare
event in the Galaxy.

\subsection{Progenitor of SN 2006X}
\label{sec:2006X} \cite{pat07} detected the CS material in SN 2006X
from variable Na I D (However, see \cite{chu08}). They found
relatively low expansion velocities (Mean velocity is about $\sim
50\ {\rm km\ s^{-1}}$), and estimated that the absorbing dust is a
few $10^{16}$ cm from the SN. A possible interpretation given by
\cite{pat07} is that the high-velocity ejecta of nova bursts in the
progenitors are, by sweeping up the stellar wind of the donor star,
slowed down to relatively low expansion velocities. According to the
resolved light echo, \cite{wan08b} fond that the illuminated dust is
$\sim$27---170 pc. The dust surrounding SN 2006X has multiple shells
\citep{wan08b}, and stems from the progenitor. In addition, the dust
surrounding SN 2006X is quite different from that observed in the
Galaxy, and has a much smaller grain size than that of typical
interstellar dust \citep{wan08a,wan08b}.

Surprisingly, there is apparent presence of dust in the CS
environment in the 2006 outburst of recurrent nova RS Oph
\citep{obr06,bod07}. \cite{eva07} and \cite{bar08} reported that the
dust appears to be present in the intervals of outbursts and is not
created during the outburst event. This means that there is a dense
region of the wind from the RG so that the dust can be formed
\citep{bod07}. According to \cite{obr06} and \cite{bod07}, the
densest parts of the red-giant wind lie in the equatorial regions
along the plane of the binary orbit. The dust in recurrent nova RS
Oph originates possibly from the dense equatorial disk around the
RG.

Therefore, it is possible that the progenitor of SN 2006X is very
similar with recurrent nova RS Oph. We assume that the candidates
for the progenitor of SN 2006X satisfy with following two
conditions: (i) there is a dense equatorial disk around RG, which
can produce dust; (ii) the progenitor of SN 2006X has undergone a
series of weak thermonuclear outbursts presented by the filled
triangles in Fig. \ref{fig:mprg} prior to explosion as SN
Ia\footnote{The thermonuclear outburst in this work has an ejecta
without an increase of CO WD's mass (strong hydrogen-shell burning),
or has an increase of CO WD's mass without an ejecta (mildly
hydrogen-shell burning). However, according to \cite{yar05}, there
is some material ejected during a mildly hydrogen-shell burning. We
assume that the progenitor of SN 2006X has undergone a mildly
hydrogen-shell burning.}. The high-velocity ejecta from every
thermonuclear outburst blow off the parts of the stellar wind
including the dust produced by dense equatorial disk. During this
process, the high-velocity ejecta are slowed down, and the dust can
be irradiated so that the grain size becomes small. A more detailed
model for SN 2006X is in preparation.

\begin{figure}
\begin{tabular}{c}
\includegraphics[totalheight=3.2in,width=3.0in,angle=-90]{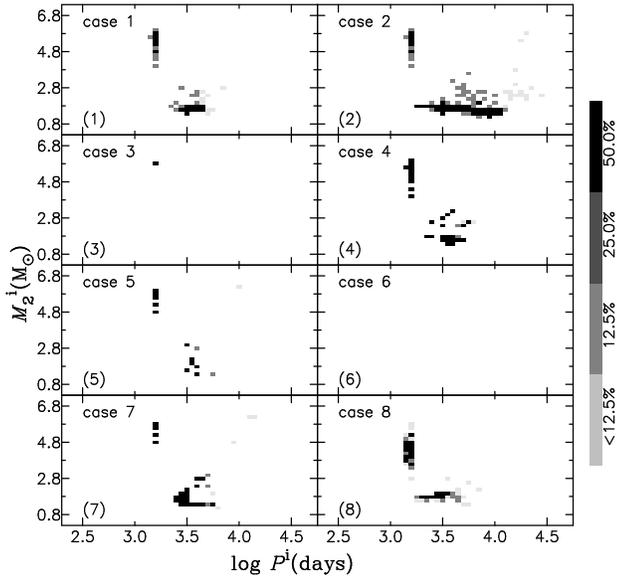}
\end{tabular}
\caption{---Similar to Fig. \ref{fig:mipi}, but for progenitors of
            SN 2006X.
            }
\label{fig:mp06}
\end{figure}

\begin{figure}
\begin{tabular}{c}
\includegraphics[totalheight=3.2in,width=3.0in,angle=-90]{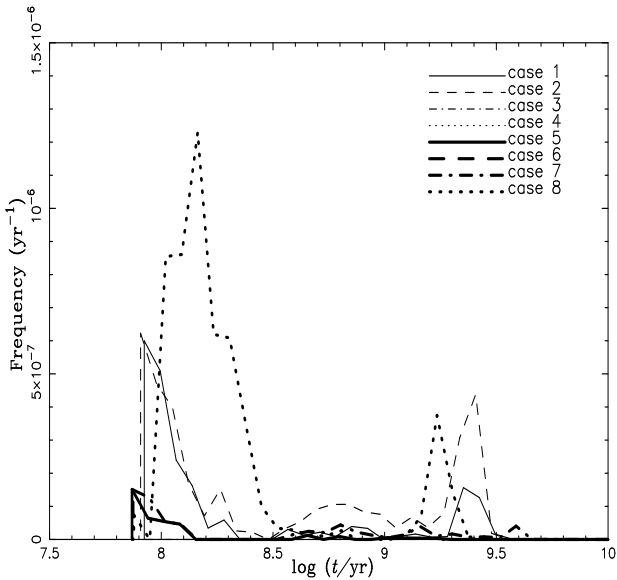}

\end{tabular}
\caption{---Similar to Fig. \ref{fig:sinb}, but for progenitors of
            SN 2006X.
            }
\label{fig:sin06}
\end{figure}

According to the above conditions, we select the possible
progenitors of SN 2006X from our sample which are showed in Fig.
\ref{fig:mp06}. The birthrates of SNe Ia like SN 2006X in the Galaxy
from cases 1 to 8 are $9.3\times 10^{-6}, 2.5\times 10^{-5},
1.0\times 10^{-7}, 6.3\times 10^{-6}, 1.2\times 10^{-6}, \sim 0.0,
3.8\times 10^{-6}$ and $7.0\times 10^{-6} {\rm yr^{-1}}$,
respectively. Therefore, SNe Ia like SN 2006X are very rare events.
Fig. \ref{fig:sin06} displays the evolution of birthrates of SNe Ia
like SN 2006X for a single star burst.



\section{Conclusion}
By a toy model in which the cool giant in symbiotic star has an
aspherical stellar wind with an equatorial disk,  we investigate the
production of SNe Ia via symbiotic channel. We estimate that the
Galactic birthrate of SNe Ia via symbiotic channel is between
$1.03\times 10^{-3}$ (case 2) and $2.27\times 10^{-5}$ yr$^{-1}$
(case 6), the delay time of SNe Ia has wide range from $\sim$ 0.07
to 5 Gyr. The results are greatly affected by the outflow velocity
and mass-loss rate of the equatorial disk. The progenitor of SN
2002ic may be a WD+RG in which there is a dense equatorial disk
around the RG. The CS environment of SN 2002ic may mainly originate
from the dense equatorial disk. In the progenitor of SN 2006X, the
CO WD has undergone a series of mildly unstable hydrogen-shell
burning, and the dust in the CS environment may originate from a
dense equatorial disk around the RG and is swept out by the ejecta
of every thermonuclear outburst.


\section*{Acknowledgments}
We are grateful to the referee, N. Soker, for careful reading of the
paper and constructive criticism. We thank Zhanwen Han for some
helpful suggestions. This work was supported by National Science
Foundation of China (Grants Nos. 10647003 and 10763001) and National
Basic Research Program of China (973 Program 2009CB824800).


\bsp

\label{lastpage}

\end{document}